# Tuning interfacial Dzyaloshinskii-Moriya interactions in thin amorphous ferrimagnetic alloys

Y. Quessab,[1,*] J.-W. Xu,[1] C. T. Ma,[2] W. Zhou,[2] G. A. Riley,[3,4] J. M. Shaw,[3] H. T. Nembach,[3,5] S. J. Poon,[2] and A. D. Kent[1]

[1]Center for Quantum Phenomena, Department of Physics, New York University, New York, New York 10003, USA
[2]Department of Physics, University of Virginia, Charlottesville, Virginia 22904, USA
[3]Quantum Electromagnetics Division, National Institute of Standards and Technology, Boulder, Colorado 80305, USA
[4]Center for Memory and Recording Research, University of California San Diego, La Jolla, CA92093, USA
[5]JILA, University of Colorado, Boulder, Colorado 80305, USA

## Abstract

Skyrmions can be stabilized in magnetic systems with broken inversion symmetry and chiral interactions, such as Dzyaloshinskii-Moriya interactions (DMI). Further, compensation of magnetic moments in ferrimagnetic materials can significantly reduce magnetic dipolar interactions, which tend to favor large skyrmions. Tuning DMI is essential to control skyrmion properties, with symmetry breaking at interfaces offering the greatest flexibility. However, in contrast to the ferromagnet case, few studies have investigated interfacial DMI in ferrimagnets. Here we present a systematic study of DMI in ferrimagnetic CoGd films by Brillouin light scattering. We demonstrate the ability to control DMI by the CoGd cap layer composition, the stack symmetry and the ferrimagnetic layer thickness. The DMI thickness dependence confirms its interfacial nature. In addition, magnetic force microscopy reveals the ability to tune DMI in a range that stabilizes sub-100 nm skyrmions at room temperature in zero field. Our work opens new paths for controlling interfacial DMI in ferrimagnets to nucleate and manipulate skyrmions.

*yassine.quessab@nyu.edu

## Introduction:

Magnetic skyrmions due to their non-trivial topology have interesting properties[1-3] that make them attractive for spintronic applications, such as racetrack memory and logic devices[4-6]. A magnetic skyrmion designates a chiral spin texture with a whirling spin configuration[7]. Skyrmions can be stabilized by broken inversion symmetry and chiral interactions, such as the Dzyaloshinskii-Moriya interactions (DMI)[8,9], which is an antisymmetric exchange interaction that favors non-collinear neighboring spins. Ultrathin magnetic materials with interfaces to heavy non-magnetic metals with large spin-orbit coupling exhibit interfacial DMI that stabilizes skyrmions and chiral domain walls[10-13]. The interfacial DMI and the nucleation of skyrmions have been extensively investigated in ferromagnetic materials[10,14-18]. Very recently, magnetic skyrmions and chiral domains were reported in ferrimagnetic systems[19-21]. Nearly compensated thin ferrimagnetic films with interfacial DMI are interesting materials due to their low stray fields, reduced sensitivity to external magnetic fields, and fast spin dynamics, which are predicted to lead to ultrasmall and ultrafast skyrmions[19,22]. Unlike in ferromagnets where fast current-induced motion of chiral textures is impeded by the Walker breakdown and domain wall pinning[13,23-25], high domain wall velocities—reaching 1000 m s$^{-1}$—have been observed in ferrimagnetic CoGd films near the angular momentum compensation temperature[19]. Hence, ferrimagnetic thin films are promising candidates for ultrafast skyrmion-based spintronics.

Recently, bulk DMI was reported in an amorphous ferrimagnetic GdFeCo alloy[26]. However, the significant advantages of interfacial DMI are that it can be controlled by the nature of the interfaces and widely tuned to stabilize skyrmions. Yet, interfacial DMI has not been studied in this class of materials. The DMI competes with the perpendicular magnetic anisotropy and Heisenberg exchange interaction and tuning the DMI in a range that favors small skyrmions can be challenging. Asymmetric domain wall nucleation and motion is commonly used to measure DMI[12,27]. However, these methods require advanced models of the domain wall dynamics[28,29]. A more direct method is Brillouin light scattering (BLS) in the Damon-Eshbach geometry, which relies on the asymmetric spin-wave frequency dispersion in the presence of DMI[30]; the asymmetry is directly related to the strength of the DMI.

Here we present a systematic study of the interfacial DMI in CoGd thin films by BLS as a function of the capping layer composition and magnetic layer thickness. We aim to understand how the DMI in CoGd films depends on the structural symmetry and magnetic properties. The interfaces are studied by cross-sectional transmission electron microscopy (TEM). We found that as little as 10% of W in the cap layer in Pt/CoGd/Pt$_{1-x}$W$_x$ thin films is sufficient to induce a DMI of about 0.15 mJ m$^{-2}$, larger than the bulk DMI found in much thicker films[26]. We also observed that the DMI is inversely proportional to the magnetic thickness in asymmetric CoGd stacks, confirming the interfacial nature of the DMI. In addition, we were able to tune the DMI in a range that stabilizes sub-100 nm skyrmions at room temperature in zero field, as observed by magnetic force microscopy (MFM). Our findings provide insight into the key parameters that control the DMI in ferrimagnetic films toward achieving ultrasmall and ultrafast skyrmions.

**Results:**

**Tuning the DMI with the capping layer composition:**

The ferrimagnetic CoGd thin films were grown by RF magnetron co-sputtering on oxidized silicon wafers in the following sequence: W(3)/Pt(3)/Co$_{78}$Gd$_{22}$(t)/Pt$_{1-x}$W$_x$(3)/Pt(3) (thicknesses in nanometers) [Methods]. The W/Pt seed layer provides good adhesion to the substrate and texture to ensure perpendicular magnetic anisotropy (PMA). The top Pt layer prevents sample oxidation. The DMI of a 5-nm thick CoGd film was studied as a function of the W composition (x) of the cap layer Pt$_{1-x}$W$_x$. The magnetic properties of the films were measured by vibrating sample magnetometry (VSM) and are summarized in Table I (Methods). Figures 1(a) and 1(b) show an out-of-plane field room-temperature magnetization hysteresis loop and the temperature (T) dependence of the saturation magnetization (M$_S$) for Pt/CoGd(5 nm)/W, respectively. M$_S$(T) is greatly reduced around 150 K, corresponding to the magnetization compensation temperature (T$_M$).

Spin wave spectroscopy using BLS was performed to measure the DMI in the CoGd films (Methods). The DMI leads to an asymmetric frequency dispersion of the counterpropagating spin waves[30]. The DMI energy $D$ (mJ m$^{-2}$) is proportional to the frequency shift ($\Delta f_{DMI}$) and given by:

$$|D| = \frac{1}{2} \left| \frac{h}{g\mu_B} \right| \frac{M_S}{k} \Delta f_{DMI}, \tag{1}$$

where $g$ is the spectroscopic splitting factor (we take $g = 2$), $\mu_B$ the Bohr magneton and $h$ Planck's constant and $k = 16.7$ $\mu$m$^{-1}$ is the spin wave vector. Notably, the DMI energy given by BLS measurements is an effective DMI averaged over the film thickness, i.e., a sum of the bottom and top interfacial contributions. An example of BLS spectra is displayed in Fig. 1(c) for Pt/CoGd(5 nm)/W. We fitted the spectra for positive and negative field polarity. The frequency shift was determined for the Stokes and the anti-Stokes peaks separately and then averaged.

The diameter of a skyrmion results from the competition between different energies such as the Heisenberg exchange energy, the magnetic anisotropy energy and the DMI strength. Ultrasmall skyrmions can be nucleated at room temperature only in a narrow range of $D$[31-33]. For DMI strength larger than a scale set by the magnetic anisotropy, the formation of stripe domains become energetically favorable[32,33]. Conversely, a weak DMI cannot stabilize a skyrmion. Theoretical work has predicted that ferrimagnetic materials are better candidates than ferromagnets to host ultrasmall and ultrafast skyrmions due to their low saturation magnetization, which causes only small stray fields[22]. Indeed, in ferrimagnetic materials, the interfacial DMI can dominate over the dipolar interactions and enable the formation of ultrasmall DMI skyrmions, which is difficult to achieve in ferromagnets. Hence, our goal was to provide a new method for controlling the interfacial DMI in thin ferrimagnetic CoGd films, which could allow one to precisely tune the DMI in a range that would enable skyrmion nucleation.

Changing the nature of the CoGd interfaces can be used to engineer the DMI strength. Therefore, the idea is to leave the Pt underlayer at the bottom interface of the CoGd film unchanged and insert a Pt$_{1-x}$W$_x$ alloy at the top interface. Thus, by changing the composition of the Pt$_{1-x}$W$_x$ alloy, the structural symmetry of the film can be gradually broken to induce DMI.

Pt is chosen for its strong spin-orbit coupling that gives a large interfacial DMI on Co spins[34,35]. Theoretical calculations based on Hund's first rule have shown that, on the contrary, a weaker DMI arises from interactions between W and Co[34] and with the same chirality as Pt and Co. In addition, W, due to its giant spin-Hall angle[36,37], would serve as a spin current source to enable skyrmion motion induced by spin-orbit torque (SOT).

Figure 2(a) shows the DMI energy as a function of the W composition (x) in Pt/CoGd(5 nm)/Pt$_{1-x}$W$_x$ measured by BLS. A maximum DMI of $0.23 \pm 0.02$ mJ m$^{-2}$ is obtained for the asymmetric stack (x = 1). In comparison, a bulk DMI of up to 0.10 mJ m$^{-2}$ was reported in ferrimagnetic GdFeCo films[26]. Conversely, the DMI is almost zero for the symmetric film (x = 0). Indeed, in the Pt/CoGd/Pt film, the top and bottom interfaces induce an interfacial DMI of similar amplitude but opposite sign (as the DMI is a chiral interaction), thus, resulting in a near vanishing effective DMI. In Pt/CoGd/W, since W gives rise to a weaker interfacial DMI, the contributions of the two interfaces are not compensated, leading to a larger effective DMI. As seen in Fig. 2(a), as little as 10 % of W introduced in the top layer is enough to significantly break the symmetry and induce a DMI of 0.15 mJ m$^{-2}$. Yet, for x > 0.1, the DMI is less sensitive to the W content in the top interface. This would indicate that for 0.1 < x < 0.9, the W rather greatly reduces the DMI between the Co spins and the Pt. Indeed, if the W were actively contributing to the interfacial DMI, a stronger dependence of the DMI energy with the alloy composition would have been expected. Additionally, the quality of the interfaces, which has a great impact on the DMI[38,39], was assessed by cross-sectional TEM. Figure 2(b) shows a cross-section of the asymmetric Pt/CoGd(5 nm)/W film, while a closer view of the top and bottom interfaces of the CoGd layer is displayed in Fig. 2(c). The CoGd alloy and the W layers are amorphous and the Pt is polycrystalline. Figure 2(c) shows that the Pt/CoGd and the CoGd/W interfaces are smooth.

**Thickness-dependence of the DMI:**

It is necessary to study the dependence of the DMI on CoGd thickness to establish its nature, i.e. to know whether the DMI is arising from interfacial effects. In our Pt/Co$_{78}$Gd$_{22}$(t)/Pt$_{1-x}$W$_x$ films, the W composition (x) was fixed either to 0 or 1 to investigate the DMI in a symmetric (x=0) and asymmetric (x=1) stack as a function of the magnetic thickness $t$. $t$ was increased from 5 nm to 15 nm. The magnetic properties were systematically measured by VSM as a function of thickness. The results are presented in Fig. 3(a) and the DMI is plotted versus the inverse magnetic thickness. In the asymmetric Pt/CoGd/W stack, the DMI is inversely proportional to the magnetic thickness and reaches a minimum of $0.09 \pm 0.01$ mJ m$^{-2}$ for $1/t = 0.067$ nm$^{-1}$ ($t = 15$ nm). The DMI linearly increases with the inverse thickness at a rate of ~ 1 mJ nm$^{-1}$. This confirms that the strength of the DMI at the interface remains unchanged and underlines its interfacial nature. In Fig. 3(b), the saturation magnetization times the magnetic thickness is plotted versus the CoGd thickness. It has a linear dependence on thickness with an x-axis intercept near zero thickness, which indicates that there is no measurable dead layer in the CoGd film. Notably, in Fig. 3(a), the intercept of the linear fit is non-zero for $1/t = 0$ (i.e. an infinitely thick film). This indicates that there is a residual DMI of 0.03 mJ m$^{-2}$, which may result from a change of the magnetization compensation temperature throughout the thickness as evidenced in another rare-earth transition-metal alloy[40] that could induce inversion symmetry breaking. Yet, as the thickness decreases, the interfacial effects become more important and the DMI increases as seen in Fig. 3(a). Thus, the interfacial DMI dominates in the entire thickness range we have studied.

On the other hand, for the symmetric Pt/CoGd/Pt film, the DMI increases with the thickness to a value of $0.09 \pm 0.02$ mJ m$^{-2}$ as seen in Fig. 3(a) (red data points). This behavior is surprising as the interfacial DMI is expected to be almost zero in symmetric layer structures. This result shows there is a difference in the nature of the top and bottom CoGd interfaces. In order to verify the latter, we performed TEM imaging in the Pt/CoGd(15 nm)/Pt film. The full stack is shown in Fig. 4(a) and a closer view of the top and bottom interfaces in Figs. 4(b) and 4(c), respectively. In Fig. 4(b), a thin layer of intermediate gray contrast (indicated by the white arrows) can be seen at the top CoGd interface and not in the bottom interface. It appears that the Pt from the capping layer has diffused into the amorphous CoGd film. As a result, the bottom and top interfaces have different roughness and intermixing. Hence, the DMI contributions of the top and bottom interfaces are not equal. Thus, due to the chirality of the interaction, they do not cancel out, leading to an increase of the net DMI. Intermixing and roughness effects appear to be more predominant in thicker films as the DMI increases with thickness in Pt/CoGd/Pt as seen in Fig. 3(a). Interestingly, the residual DMI observed for the asymmetric stack is of the same order of magnitude as that for the symmetric Pt/CoGd(5 nm)/Pt structure. Therefore, this suggests that this residual DMI is independent of the CoGd interfaces. Finally, as the skyrmion size depends on the magnetic film thickness[22,33], it is thus important to understand how the interfacial DMI scales with the thickness.

**Evidence of magnetic skyrmions by MFM:**

Finally, we aimed to verify whether these thin ferrimagnetic alloy films would indeed host skyrmions. We focused on the asymmetric Pt/CoGd/W stacks as they are more promising for skyrmion motion via spin-orbit torque because of the giant spin-Hall angle of W[36,37]. In fact, in Pt/CoGd/Pt, the spin-orbit torques from the top and bottom interfaces would tend to cancel each other out. The Pt/CoGd/W films were subject to AC in-plane magnetic field demagnetization and imaged by atomic and magnetic force microscopy (AFM and MFM) at room temperature in zero field. Figure 5 shows images for Pt/CoGd(10 nm)/W. The left column is AFM data (Figs. 5(a) and (c)) and right column MFM images (Figs. 5(b) and (d)). The surface roughness was measured and is on the order of 0.2 nm (rms). Magnetic contrast is indicated by dark areas in the MFM images. By comparing the AFM and MFM images, it is clear that this contrast comes from magnetic textures and is not due to topography. Several skyrmion-like textures can be seen in Fig. 5(b). Figure 5(d) corresponds to a smaller MFM scan performed around of one of them marked by a square box in Fig. 5(b). This skyrmion-like texture is on the order of 100 nm. 50 nm skyrmions were observed in Pt/CoGd(8 nm)/W (see supplemental materials[41]). Arguably, considering the size of these textures, the DMI energy values [see Fig. 3(a)], and the fact that the CoGd films are weakly magnetized ($M_S \sim 140 - 150$ kA m$^{-1}$, see supplemental materials[41]), it is unlikely that these textures be magnetic bubbles[4,7] stabilized by dipolar interactions. Thus, MFM images would rather indicate the presence of skyrmions. However, accurate estimation of the skyrmion size is difficult. Indeed, the MFM tip is sensitive to the dipolar field emerging from the magnetic texture which is spatially spread out. Furthermore, smaller magnetic features may be present in Fig. 5(b), yet they cannot be clearly distinguished due to the background noise and small magnetic contrast.

**Discussion:**

To summarize, we have demonstrated that by capping the ferrimagnetic CoGd layer with a PtW alloy we could tune the DMI energy over a large range, from almost no DMI to an interfacial DMI energy of 0.23 mJ m$^{-2}$. The DMI thickness dependence reveals the interfacial nature of the DMI in CoGd thin films. Thus, the DMI strength can be controlled by the interfaces in the thickness range we studied, which is also the range relevant for skyrmion nucleation. Moreover, the DMI was found to be non-zero in thicker symmetric structures emphasizing the role of interface roughness and intermixing. Lastly, we showed evidence that films can have a DMI in a range that allows sub-100 nm skyrmion nucleation at room temperature in zero field. Our experimental results provide insight into the key parameters that control the DMI in ferrimagnetic films toward achieving ultrasmall and ultrafast skyrmion motion for spintronic applications.

**Methods:**

**Thin film deposition:**

The thin films were prepared by RF magnetron sputtering and deposited onto Si-SiO$_2$ substrates at room temperature with a base pressure of $2.7\times10^{-5}$ Pa. The Ar deposition pressures of W, Pt, CoGd, and Pt$_{1-x}$W$_x$ were 0.93 Pa, 0.1 Pa, 0.16 Pa, and 0.16 Pa, respectively. CoGd films were obtained by co-sputtering from the Co and Gd targets. The powers of the Co and Gd sources were tuned to obtain CoGd films with approximately 78 at. % of Co. The deposition rates were calibrated using x-ray reflectometry.

**Magnetometry:**

The magnetic properties of the samples were measured by vibrating sample magnetometry. Magnetic hysteresis loops were measured by varying the temperature from 100 K to 300 K with steps of 25 K in order to extract the temperature dependence of the saturation magnetization and the coercive field. Magnetometry was systematically performed prior to BLS experiments.

**Brillouin light scattering:**

Spin wave spectroscopy using BLS is sensitive to interfacial effects and can be used to measure the DMI strength. The spin waves (SWs) inelastically scatter the monochromatic laser beam that is focused onto the sample surface. The frequency of the scattered photons is shifted by the SWs frequency. The SWs frequency is determined by analyzing the backscattered light with a (3 + 3)-pass tandem Fabry-Pérot interferometer. The counterpropagating Damon-Eshbach SWs have a non-reciprocal frequency dispersion characterized by a frequency shift (noted $\Delta f_{DMI}$). The frequency shift is considered here in absolute value. An in-plane bias magnetic field was applied to allow the SW propagate in-plane (Damon- Eshbach geometry). For a $\lambda = 532$ nm laser beam with an incidence of $\theta_i = \pi/4$, the SW vector, $k$ defined as $k = 4\pi\sin(\theta_i)/\lambda$ was set to 16.7 μm$^{-1}$.

**Acknowledgments:**

This work was supported by DARPA grants No. D18AP00009 and R186870004 and by the DOE grant No DE-SC0018237.

**Author contributions:**

A.D.K. and S.J.P. conceived of and supervised the project. Y.Q, H.T N., S.J.P. and A.D.K. planned the experiments. BLS measurements were conducted by G.A.R., J.M.S. and H.T.N. BLS data was analyzed by Y.Q., J.-W.X. A.D.K. and H.T.N. Sample fabrication and magnetometry measurements were performed by C.T M. and W.Z. MFM imaging was done by

| x | $M_S$ (kA m$^{-1}$) | $\mu_0 H_C$ (mT) | $T_M$ (K) |
|---|---|---|---|
| 0 | 160 | 12.5 | 125 – 150 |
| 0.1 | 166 | 11.0 | 125 |
| 0.25 | 160 | 10.0 | 125 |
| 0.5 | 180 | 12.0 | 125 – 150 |
| 0.75 | 160 | 12.0 | 125 – 150 |
| 1 | 145 | 11.0 | 125 – 150 |

TABLE I. Summary of the magnetic properties of Pt/Co$_{78}$Gd$_{22}$(5 nm)/Pt$_{1-x}$W$_x$ films as a function of W composition (x). The room temperature saturation magnetization ($M_S$) and coercive field ($\mu_0 H_C$) are indicated.

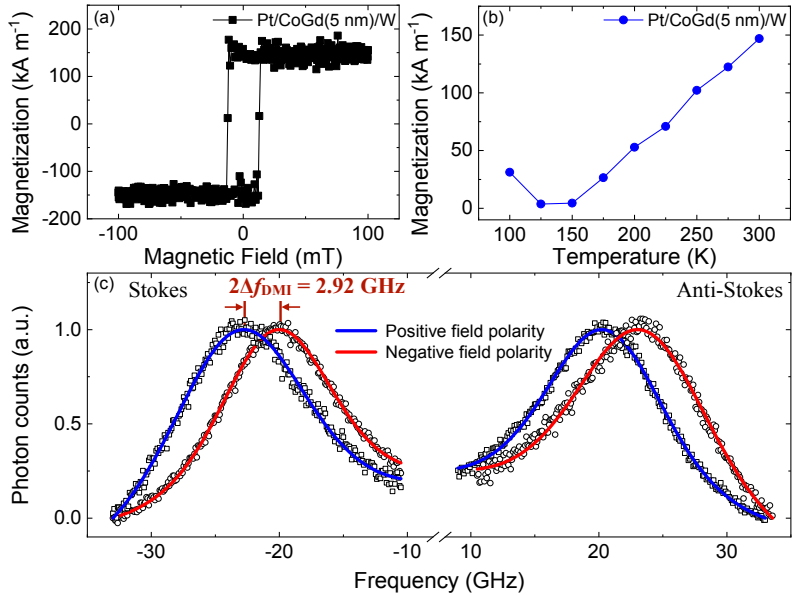

FIG. 1. (a) Out-of-plane magnetization hysteresis loop (b) and temperature dependence of the saturation magnetization measured by VSM for the Pt/CoGd(5 nm)/W sample. Magnetic compensation of this CoGd composition occurs around 150 K. (c) Spin wave spectroscopy obtained by BLS in Pt/CoGd(5 nm)/W. The shift in the frequency dispersion, $\Delta f_{DMI}$, is proportional to the DMI. The applied in-plane field was 0.460 T. The solid lines are fit to the BLS data obtained for positive (blue curve) and negative (red curve) field polarity.

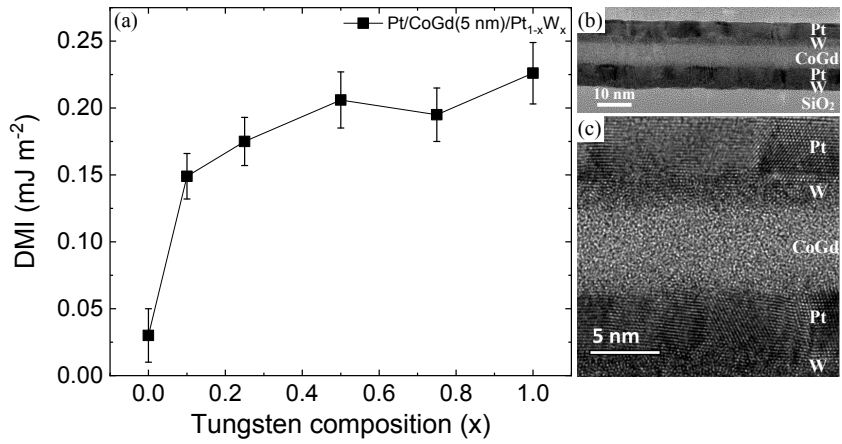

FIG. 2. (a) DMI energy measured by BLS in Pt/CoGd(5 nm)/Pt$_{1-X}$W$_{X}$ as a function of the W composition (x). The solid black line is a guide to the eye. (b) and (c) Cross-sectional TEM images of the Pt/CoGd(5 nm)/W film. (b) The full stack and (c) a magnified view of the top and bottom interface of the CoGd layer.

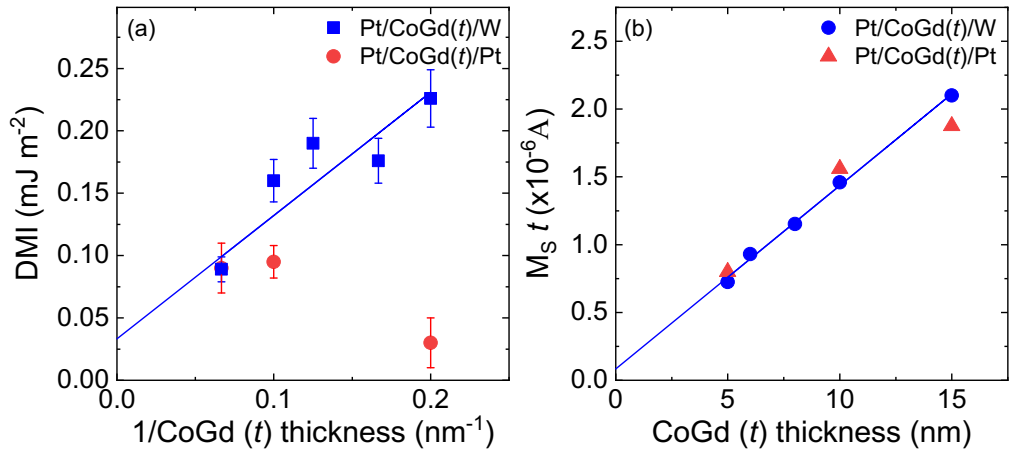

FIG. 3. (a) Magnetic thickness (*t*) dependence of the DMI in Pt/CoGd(t)/(W or Pt) with the DMI energy plotted against 1/*t*. In Pt/CoGd/W, the increase indicates the interfacial nature of the DMI interactions. (b) Room temperature magnetization thickness product versus thickness. The solid blue lines are linear fits to the Pt/CoGd/W data.

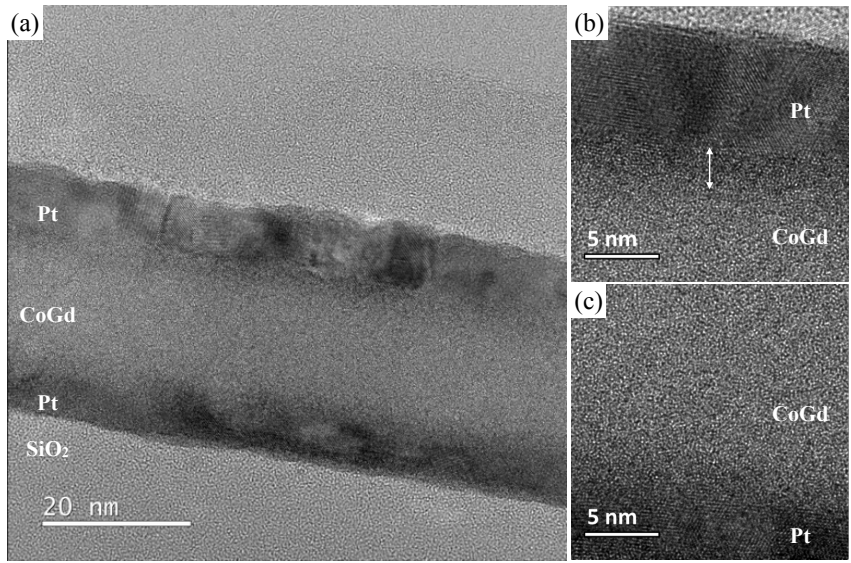

FIG. 4. Cross-sectional TEM images of the symmetric Pt/CoGd(15 nm)/Pt structure. The full stack is shown in (a) with a magnified view of the top (b) and bottom (c) interface of the CoGd layer.

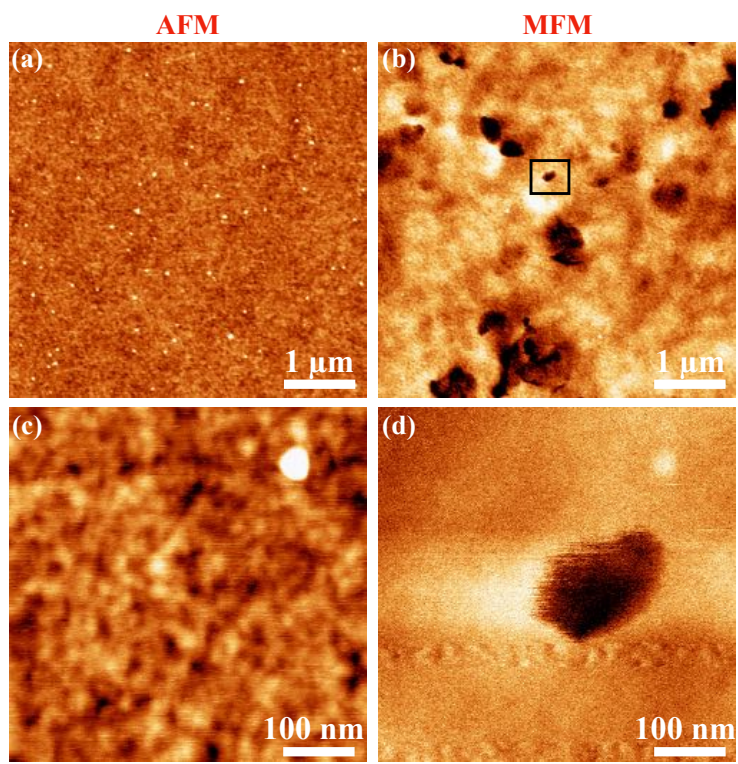

FIG. 5. AFM (a,c) and MFM (b,d) images showing skyrmion-like magnetic textures nucleated in Pt/CoGd(10 nm)/W at room temperature in zero-field. The skyrmion imaged in (d) is indicated by a square box in (b).